\documentclass[a4paper,11pt]{article}

\usepackage{pos}
\usepackage{lipsum} 
\usepackage{wrapfig}
\usepackage{natbib}

\title{Operations plans and sensitivities of the IceCube Upgrade Camera System}

\author{The IceCube Collaboration \\{\normalsize \normalfont(a complete list of authors can be found at the end of the proceedings)}\\}

\emailAdd{srodan@icecube.wisc.edu, rott@physics.utah.edu}

\abstract{
The IceCube Upgrade consists of seven new strings to be deployed in the central region of the existing IceCube detector.  The goals of the IceCube Upgrade are two-fold: to enhance sensitivity to neutrinos in the GeV range, and to improve the calibration of the IceCube detector as a means of reducing systematic uncertainties due to the optical properties of the ice.
Among other calibration devices designed to study ice properties, a novel camera system will be deployed as part of the Upgrade. The system will include three cameras, each paired with an illumination LED, included in each of the Upgrade optical modules.  In total, 2,300 cameras will be deployed. A combination of photographic images from transmitted and reflected light will measure optical properties of both the bulk ice in-between strings and the local ice refrozen in the drill hole. In this contribution, we present the operations plans for these two types of measurements and the sensitivities to the ice properties and geometry of the new modules that can be achieved with the new camera system.

\vspace{4mm}
{\bfseries Corresponding authors:}
Steven Rodan$^{1*}$, Christoph T\"onnis$^{1,2}$, Jiwoong Lee$^{1}$, Carsten Rott$^{1,2}$\\
{$^{1}$ \itshape Department of Physics, Sungkyunkwan University, Suwon 16419, Rep. of Korea}\\
{$^{2}$ \itshape Department of Physics and Astronomy, University of Utah, Salt Lake City, UT 84112, USA}\\
\\[4mm]

$^*$ Presenter

\ConferenceLogo{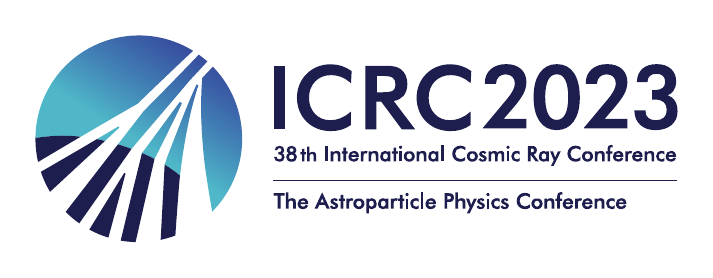}

\FullConference{The 38th International Cosmic Ray Conference (ICRC2023)\\ 26 July -- 3 August, 2023\\ Nagoya, Japan}
}

\begin{document}

\maketitle

\section{Introduction}\label{sec1}

Located at the geographic South Pole, the IceCube Neutrino Observatory \cite{Aartsen:2016nxy} is an optical Cherenkov detector, encompassing a cubic kilometer of ultra-pure Antarctic ice found at depths ranging from 1.45 km to 2.45 km.
Additionally, IceCube features an air-shower detector spanning one square kilometer on the surface of the ice \cite{icetop}. The primary objectives of this observatory are to measure high-energy astrophysical neutrino fluxes and ascertain their sources.

\begin{figure}[h]
    \centering
    \includegraphics[width=\textwidth]{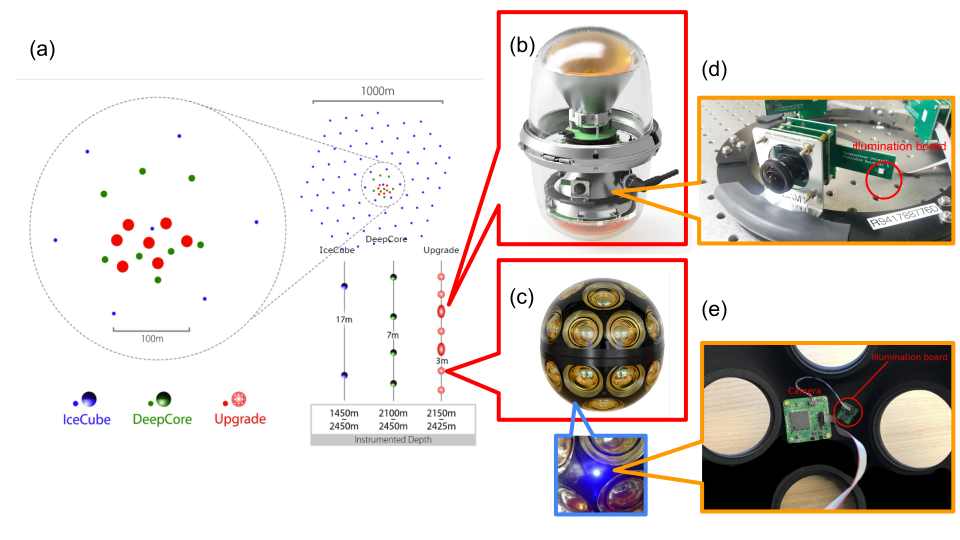}
    \caption{(a) A diagram showing where the Upgrade strings (red) will be deployed with respect to current IceCube (blue) and DeepCore (green). A comparison of the vertical spacing for the three subdetectors is also depicted. (b) and (c) show the D-Egg and mDOM modules, respectively. (d) and (e) show zoomed-in views of how the camera system is integrated in the D-Egg and mDOM.}
    \label{fig:intro}
\end{figure}

The IceCube Upgrade \cite{Ishihara:2019uL} consists of seven densely instrumented strings in the central region of the active volume of IceCube. It incorporates digital optical modules (DOMs) developed with new designs and updated electronics. DOMs on each string are vertically spaced every 3 meters, between depths of 2160 meters and 2430 meters beneath the ice's surface, as depicted in Fig.~\ref{fig:intro}(a). The majority of DOMs to be deployed are of two types:
the "D-Egg" with two 8-inch photomultiplier tubes (PMTs), one oriented upwards and the other downwards (see Fig.~\ref{fig:intro}(b)), and the "mDOM", equipped with 24 three-inch PMTs distributed for near-uniform, isotropic coverage (see Fig.~\ref{fig:intro}(c)). 

A significant objective of the IceCube Upgrade is to improve the accuracy of the optical properties of the detector medium, both the undisturbed "bulk" ice and re-frozen "hole" ice, thereby reducing uncertainties in neutrino energy and angular reconstruction. The scattering and absorption lengths of the ice within the IceCube detector have mainly been probed by injecting photons into the ice using LEDs included in IceCube DOMs and recording the pattern of light observed in nearby DOMs \cite{IceCal1}. More recently, an optical anisotropy due to birefringence in the ice was discovered using this system \cite{Abbasi:2021q6}. Despite these milestones in understanding the detector medium, there remain systematic uncertainties \cite{IceCal1}, including notably the column of bubbles and other defects that form in the hole ice, for which the novel camera system will play a crucial role in measuring.

The camera system will complement other calibration devices towards better understanding the bulk ice and lead to an improved ice model that will enhance the accuracy of over 17 years of IceCube data when retroactively applied.

\section{The IceCube Upgrade Camera System}\label{sec2}

The IceCube Upgrade Camera System is the same in all DOM types. It uses a Sony IMX225LQR-C color CMOS image sensor, with a 1312 by 993 pixel resolution, and has a power consumption of 2.3~W. It will be capturing images of light produced by an OSLON SSL series light emitting diode (LED) on an illumination board. The 465-nm wavelength LED light cone has an opening angle of 80$^{\circ}$ and uses 1.2~W of power. Images will be taken of the LED light with an exposure time of up to one minute, and an analog gain of up to 30 decibels.
Depending on the image from a LED at a known distance and orientation, one can ascertain the optical properties of the medium. 

In the D-Egg, three cameras are equally-spaced on a ring installed in the lower half. An illumination board with one LED is mounted next to the cameras, facing the same direction. In the mDOM, two cameras are angled at 45$^{\circ}$ in the upper hemisphere, facing opposite directions, while a third camera is installed at the bottom, pointing downward. Additionally, there is a separate illumination board installed at the top, facing upward (see Fig.~\ref{fig:intro}(e)). Further details regarding the camera system hardware can be found in \cite{IceCube:2021jfx}.

\section{Geometry of the IceCube Upgrade}\label{sec3}

\begin{figure}[h]
    \centering
    \includegraphics[width=\textwidth]{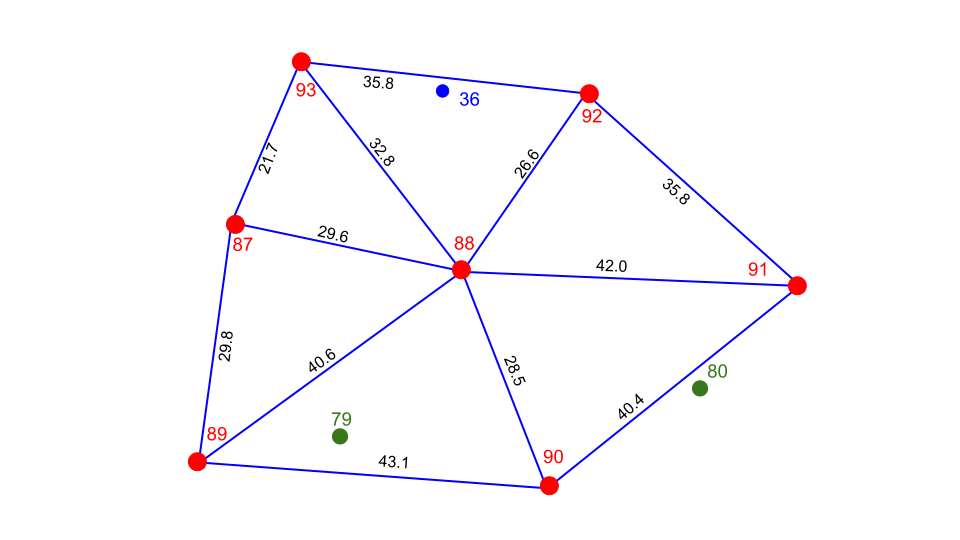}
    \caption{A diagram of the IceCube Upgrade geometry, as currently envisioned.
    The seven Upgrade strings are numbered 87 through 93.  IceCube string 36
    and IceCube DeepCore strings 79 and 80 fall within this footprint, and these
    existing strings are indicated by the blue and green circles, respectively.
    Distances between the strings are shown in meters.}
    \label{fig:surface_geom}
\end{figure}

The currently envisioned string layout for the Upgrade is shown in Fig.~\ref{fig:surface_geom}, including the string numbers, all inter-string distances (in meters), as well as the locations of a few deployed strings that will fall within the Upgrade volume. The horizontal spacing between strings typically will range from 21~m to 43~m, and the vertical spacing between DOMs on the same string will be 3~m.

The vertical variation in inter-DOM distance is conservatively expected to be on the order of 15~cm. Due to the unavoidable deviation of the bore hole during drilling, the sideways variation of DOM locations is also expected to be closer to 1.5~m. This may be negligible for bulk ice measurements, where the distance between strings are orders of magnitude larger, but, for the hole ice, these variations must be measured locally between nearby DOMs with the aid of simulations.

\section{Operations Plan}\label{sec4}

The camera system was designed to be used in several different measurements, which include, in order of priority: hole ice, geometry and bulk ice measurements. For each proposed measurement an accompanying script will be written for automated execution by scientists at the South Pole. 

The camera operations need to consider power limitations and storage space in the DOMs. To prevent overload, each DOM will not operate more than one LED and one camera simultaneously. Images captured by the cameras will be transferred to the mainboard as they are captured.

DOMs are connected to the surface DAQ by twisted wire pairs. Three DOMs are connected to each wire pair and share approximately 1.5 Mbaud bandwidth to the surface. Thus, to send a single (full resolution) image of 2.7~MB would take at least ~14.4 seconds.

The exposure time must be adjusted in order to maximally utilize the dynamic
range of the camera and avoid saturation. Based on the camera's previous deployment in the SPICEcore hole \cite{Toennis:SPICECAM2021} - a drilled hole just outside of IceCube, filled in with antifreeze - exposure times of 3 to 60 seconds are expected to achieve this, depending on the LED orientation relative to the camera. For all measurements, PMTs in the local DOMs near the camera are anticipated to be switched off while operating the LEDs.

\subsection{Hole Ice Measurements}

After deploying a string of DOMs, close observation of hole ice formation becomes possible. This includes the emergence of a cluster of trapped gas bubbles near the center of the hole during the final freeze-in stages, known as the 'bubble column.' This is a result of incomplete degassing of the drill water, and while degassing is planned for the Upgrade deployment, a complete absence of bubbles cannot be expected.

\begin{figure}
    \centering
    \includegraphics[width=0.9\textwidth]{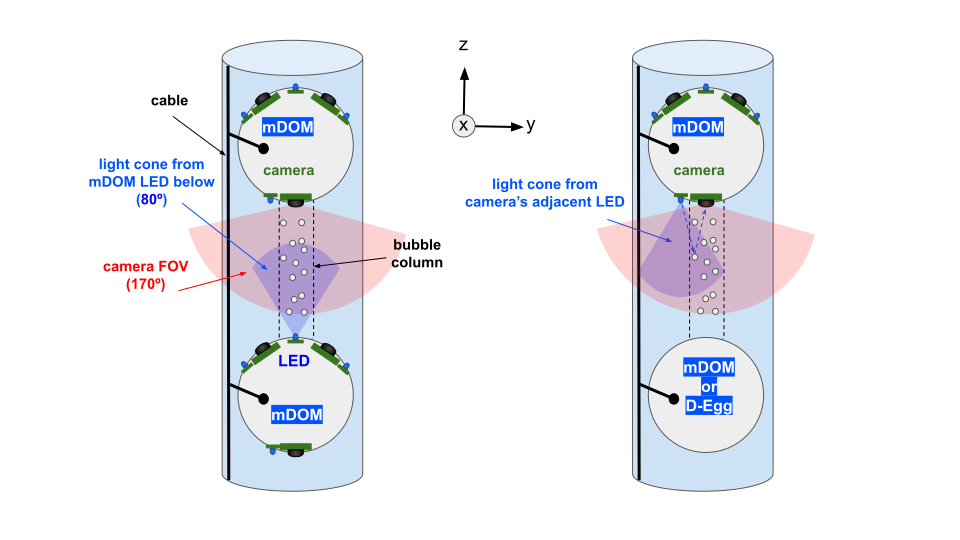}
    \caption{Diagrams of the two types of hole ice measurements, seen from the side. Downward facing mDOM cameras will capture images (left) using upward facing LED on mDOM directly below, and (right) using backscattered photons from the adjacent LED.}
    \label{fig:holeice}
\end{figure}

Images of the refrozen ice within the holes will be taken using the downward-facing camera of the mDOMs, illuminated by a single LED. In the case where another mDOM is directly below, the upward-facing LED at the top of the mDOM directly below it will be used (Fig.~\ref{fig:holeice}(L)), and irrespective of the DOM type immediately below the camera, the LED adjacent to the camera can be used (Fig.~\ref{fig:holeice}(R)). Parameters such as the bubble column diameter and its position relative to the optical module center will be extracted from these images.

In order to detect more subtle features in the hole ice, such as the hole-bulk ice transition, dust on the DOM's surface, or any fracture planes in the ice, we will use an exposure time that enables the cameras to capture enough backscattered light. In the SPICEcore 
this light was observed at 200 counts per pixel with a 3.7-second exposure time. Therefore, we anticipate that by extending the exposure time to 55 seconds would be enough to detect this light near the optimal 3000 pixel counts.

We plan to capture a total of 2-4 images per camera (making 800-1600 images in total across all 400+ mDOMs) for the hole ice measurement: two with the LED directly adjacent to the capturing camera using two different exposure times, and, in the case where another mDOM is below the camera, an additional two images will be taken with the upward-facing LED illuminated.

\subsection{Geometry Measurement}

We will briefly measure the camera positions and orientations for the subsequent longer bulk ice measurement and check for obstructions in the ice utilizing the outward-facing D-Egg cameras and the tilted upper mDOM cameras, which, for the most part, will be the same DOM type within a depth layer. We expect to be able to differentiate the LED light cones in the images, and from the resulting pixel count distribution, the relative orientation of the LED can be determined. Following this procedure, we will be able to further use triangulation to obtain the LED positions using two or more cameras. 

The sensitivity of the cameras images for extracting the LED positions can be estimated. Using the camera's reduced effective field of view (taking into account the refraction of light passing through ice, glass and air), the glass curvature and pixel resolution, this corresponds to an angular uncertainty of roughly 0.07$^{\circ}$ per pixel of uncertainty. Taking the images from two cameras on neighboring strings and facing the same LED light cone, one cane use triangulation to find the LED positions. Knowing the positions of the cameras, and with the relative orientation extracted from both images, one can find the intersection of lines of sight pointing from each camera to the LED, and so calculate the LED's position. Using error propagation, and assuming an uncertainty of just one pixel in the image's center of mass, this leads to an uncertainty in the LED position of $\sim$4~cm.

For each camera, we plan to sequentially activate LEDs on visible DOMs and record images.  Exposure time will be optimized based on the observed brightness and saturation and will be adjusted in subsequent images.

It is necessary to differentiate the light cones produced by LEDs on different DOMs. Our previous tests in a 2-m deep swimming pool  \cite{Kang:ICUcam2019} have shown that we could clearly distinguish LEDs that are 40~cm apart at a 25~m distance in water. However, given the increased scattering in ice, and with similar vertical device separation of 3~m, we prefer to operate LEDs simultaneously only if the LEDs have a separation of 30~m or more.

Taking all of these factors into consideration, the run plan for this measurement is divided into three phases, summarized in table~\ref{tab:phases}. 

\begin{figure}[h]
    \centering
    \includegraphics[width=\textwidth]{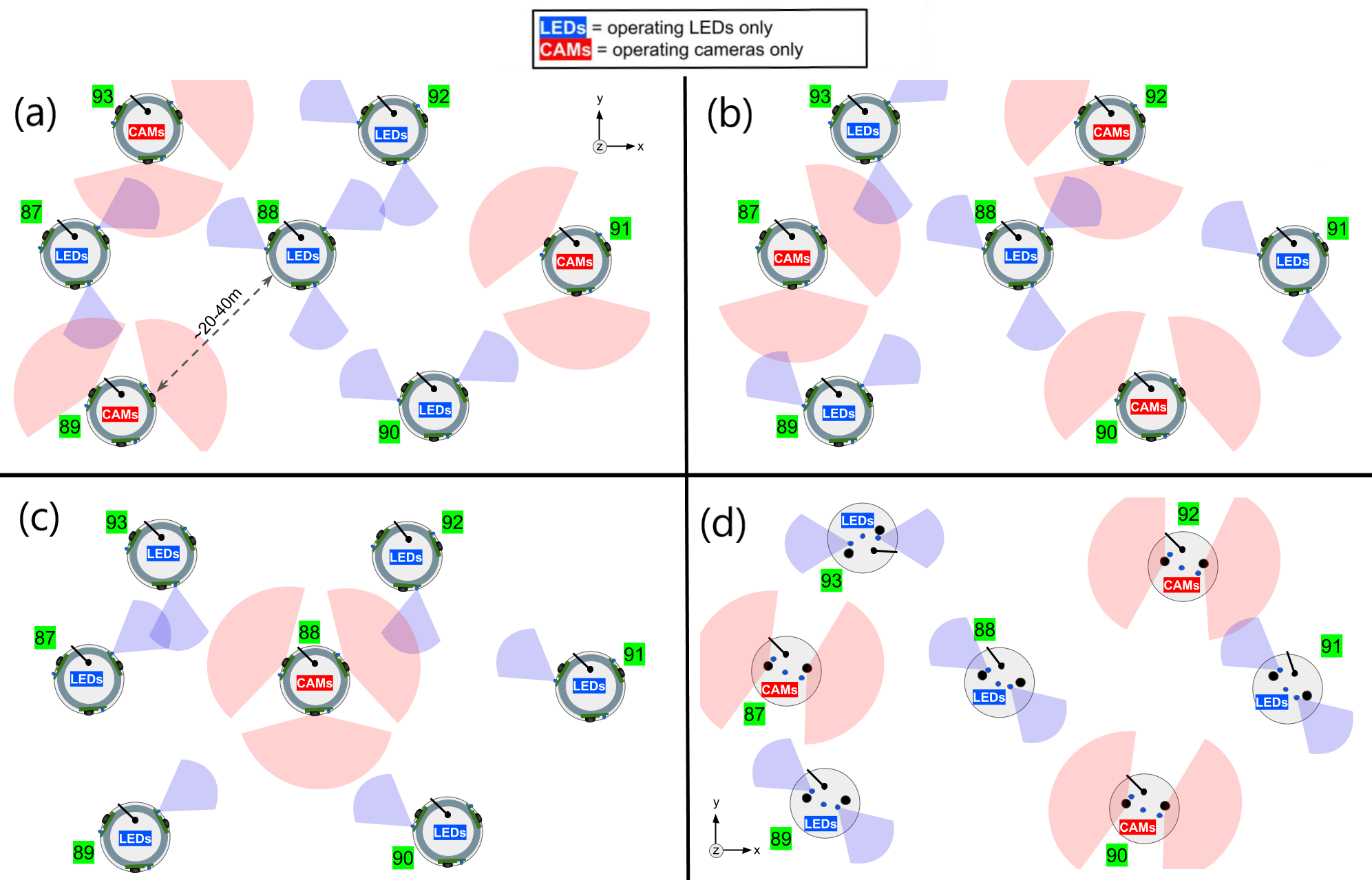}
    \caption{Schematic of geometry and bulk ice measurement phases: (a), (b) and (c) correspond to the first, second and third phases mentioned in the text, respectively, in a D-Egg layer and (d) shows the second phase in an mDOM layer, where there are two cameras in the upper hemisphere, pointing in opposite directions at a 45$^{\circ}$ above horizontal. Note: DOMs are not drawn to scale, but enlarged for visibility.}
    \label{fig:bulk_meas}
\end{figure}

During Phase 1 (Fig.~\ref{fig:bulk_meas}(a)), D-Egg cameras on strings with operating cameras will be operated one at a time. For each camera, the D-Eggs on strings with operating LEDs will illuminate their LEDs one by one. Then the second camera on the D-Eggs with operating cameras will capture images of all 3 LEDs again, and once more for the 3rd camera totalling 9 images of all camera-LED combinations.  

In Phase 2 (Fig.~\ref{fig:bulk_meas}(b)), the strings which were operating LEDs during Phase 1 will be operating cameras and vice versa, but the image capturing sequence will otherwise be the same. The central cable (string 88) will be operating LEDs during both the first and second phases, then in Phase 3 (Fig.~\ref{fig:bulk_meas}(c)), it will be operating cameras while all outer 6 strings will be operating LEDs.

The outer-string DOMs will have images from one camera that contain no information as they faced outwards and thus couldn't capture any LED light cone. Since we don't have an \textit{a priori} knowledge of which cameras these will be, after the geometry run we can discard those images. Moreover this allows us to identify those cameras that see no light cone, and shorten subsequent bulk ice measurement time by taking images with just 2/3 of the cameras on the outer six strings.

\begin{table}[h]
    \centering
    \caption {Phases of Geometry and Bulk Ice Measurements} \label{tab:phases} 
    \begin{tabular}{|c|l|l|}
        \hline
        \textbf{Phase No.} & \textbf{Strings operating cameras} & \textbf{Strings operating LEDs} \\
        \hline
        \textbf{1} & 89, 93, 91 & 87, 90, 92, 88 \\
        \hline
        \textbf{2} & 87, 90, 92 & 89, 93, 91, 88 \\
        \hline
        \textbf{3} & 88 & 87, 89, 90, 91, 92, 93 \\
        \hline
    \end{tabular}
    \label{bulk-geom_phases}
\end{table}

The same measurement scheme can be applied to the mDOM layers (Fig.~\ref{fig:bulk_meas}(d)). Two of the three mDOM cameras, positioned 45$^{\circ}$ away from the vertical direction and facing in opposite directions, will be used. Unlike the D-Egg cameras, the mDOM cameras have an azimuthal separation of 180$^{\circ}$ (rather than 120$^{\circ}$ of the former).

\subsection{Bulk Ice Measurement}

The camera system in the D-Egg modules and the top mDOM cameras tilted at 45$^{\circ}$ above horizontal are expected to measure the scattering length in the local ice with a precision of about 10 meters. This measurement relies on how accurate the orientation of a LED on a module on an adjacent string relative to the camera is determined through the geometry run (as explained above). The camera-LED combinations for capturing images will be determined in advance by analyzing the geometry run image data set.

The plan is to capture a series of bulk ice images in three stages, essentially equivalent to those of the geometry measurement section and shown in Fig.~\ref{fig:bulk_meas}.  To obtain relatively accurate scattering lengths, we will use a longer exposure time than in the geometry run so that fainter features can be resolved to differentiate images. 

After completing all three phases, 2-4 images will have been taken by the DOMs of the peripheral six strings, while DOMs on the central string will have taken 3 images. With an average exposure time of 30 seconds, and a total of 7 - 11 image capture configurations, the entire capture process is expected to take 4 - 6 minutes per DOM level. With the transfer time of one minute per 4 images, add on another 3 minutes per DOM level for a total of 7 - 9 minutes.

Like the geometry run, this measurement will run on one DOM layer per Quad at a time in order to avoid light pollution from above or below. With up to 6 DOMs per Quad, the total time required to record and transfer the bulk ice images is expected to be 42 - 54~minutes. By compressing the image bundles beforehand, the transfer time could be reduced  by a factor of $\sim$3.

\section{Conclusion and Outlook}\label{sec6}

With the completion of the IceCube Upgrade camera system production, detailed plans have been developed to operate the cameras in the ice and conduct  measurements. The purpose is to obtain images for analyzing the hole and bulk ice while minimizing the downtime of IceCube PMTs. Ongoing work includes image simulation using the Photon Propagation Code~\cite{ppc}, where template images of LED light illuminating hole and bulk ice under various conditions (e.g. various, realistic scattering lengths, camera-LED orientations, etc.) are being created. These simulated images will be used in analyses of images captured by the camera system \cite{toennis:2023icrc}.

\bibliographystyle{ICRC}
\bibliography{references}

\providecommand{\href}[2]{#2}\begingroup\raggedright\begin{thebibliography}{10}

\bibitem{Aartsen:2016nxy}
{\bfseries IceCube} Collaboration, M.~G. Aartsen {\em et~al.}
  \href{http://dx.doi.org/10.1088/1748-0221/12/03/P03012}{{\em JINST}
  {\bfseries 12} no.~03, (2017) P03012}.

\bibitem{icetop}
{\bfseries IceCube} Collaboration, R.~Abbasi {\em et~al.}
  \href{http://dx.doi.org/10.1016/j.nima.2012.10.067}{{\em Nuclear Instruments
  and Methods in Physics Research Section A: Accelerators, Spectrometers,
  Detectors and Associated Equipment} {\bfseries 700} (2013) 188--220}.

\bibitem{Ishihara:2019uL}
{\bfseries IceCube} Collaboration, A.~Ishihara {\em et~al.}
  \href{http://dx.doi.org/10.22323/1.358.1031}{{\em PoS} {\bfseries ICRC2019}
  (2019) 1031}.

\bibitem{IceCal1}
{\textbf{IceCube} Collaboration, M. G. Aarsen \textit{et al.}}
  \href{http://dx.doi.org/10.1016/j.nima.2013.01.054}{{\em Nuclear Instruments
  and Methods in Physics Research Section A: Accelerators, Spectrometers,
  Detectors and Associated Equipment} {\bfseries 711} (May, 2013) 73--89}.

\bibitem{Abbasi:2021q6}
{\bfseries IceCube} Collaboration, D.~Chirkin, M.~Rongen, {\em et~al.}
  \href{http://dx.doi.org/https://doi.org/10.22323/1.395.1119}{{\em POS}
  {\bfseries 395} no.~1119, (2021) }.

\bibitem{IceCube:2021jfx}
{\bfseries IceCube} Collaboration, W.~Kang, J.~Lee, G.~Roellinghoff, C.~Rott,
  and C.~T\"onnis \href{http://dx.doi.org/10.22323/1.395.1064}{{\em PoS}
  {\bfseries ICRC2021} (2021) 1064}.

\bibitem{Toennis:SPICECAM2021}
{\bfseries IceCube} Collaboration, C.~T\"onnis, D.~Kim, A.~Pollmann, C.~Rott,
  {\em et~al.} \href{http://dx.doi.org/10.22323/1.395.1047}{{\em PoS}
  {\bfseries ICRC2021} (2021) 1047}.

\bibitem{Kang:ICUcam2019}
{\bfseries IceCube} Collaboration, W.~Kang, C.~T\"onnis, C.~Rott, {\em et~al.}
  \href{http://dx.doi.org/10.22323/1.358.0928}{{\em PoS} {\bfseries ICRC2019}
  (2019) 928}.

\bibitem{ppc}
{\bfseries IceCube} Collaboration, {D. Chirkin}
  \href{http://dx.doi.org/10.1016/j.nima.2012.11.170}{{\em Nucl. Instrum. Meth.
  A} {\bfseries 725} (2013) 141--143}.

\bibitem{toennis:2023icrc}
{\bfseries IceCube} Collaboration, {Christoph T\"onnis, Seowon Choi, Jiwoong
  Lee, Carsten Rott} {\em PoS} {\bfseries ICRC2023} (these proceedings) 1071.

\end{thebibliography}\endgroup

\clearpage

\section*{Full Author List: IceCube Collaboration}

\scriptsize
\noindent
R. Abbasi$^{17}$,
M. Ackermann$^{63}$,
J. Adams$^{18}$,
S. K. Agarwalla$^{40,\: 64}$,
J. A. Aguilar$^{12}$,
M. Ahlers$^{22}$,
J.M. Alameddine$^{23}$,
N. M. Amin$^{44}$,
K. Andeen$^{42}$,
G. Anton$^{26}$
% ...
\\
\\
% ...
$^{12}$ Universit{\'e} Libre de Bruxelles, Science Faculty CP230, B-1050 Brussels, Belgium \\
% ...
$^{17}$ Department of Physics, Loyola University Chicago, Chicago, IL 60660, USA \\
$^{18}$ Dept. of Physics and Astronomy, University of Canterbury, Private Bag 4800, Christchurch, New Zealand \\
% ...
$^{22}$ Niels Bohr Institute, University of Copenhagen, DK-2100 Copenhagen, Denmark \\
$^{23}$ Dept. of Physics, TU Dortmund University, D-44221 Dortmund, Germany \\
% ...
$^{26}$ Erlangen Centre for Astroparticle Physics, Friedrich-Alexander-Universit{\"a}t Erlangen-N{\"u}rnberg, D-91058 Erlangen, Germany \\
% ...
$^{40}$ Dept. of Physics and Wisconsin IceCube Particle Astrophysics Center, University of Wisconsin{\textendash}Madison, Madison, WI 53706, USA \\
% ...
$^{42}$ Department of Physics, Marquette University, Milwaukee, WI, 53201, USA \\
% ...
$^{44}$ Bartol Research Institute and Dept. of Physics and Astronomy, University of Delaware, Newark, DE 19716, USA \\
% ...
$^{63}$ Deutsches Elektronen-Synchrotron DESY, Platanenallee 6, 15738 Zeuthen, Germany  \\
$^{64}$ Institute of Physics, Sachivalaya Marg, Sainik School Post, Bhubaneswar 751005, India \\
% ...

\subsection*{Acknowledgements}

\noindent
The authors gratefully acknowledge the support from the following agencies and institutions:
USA {\textendash} U.S. National Science Foundation-Office of Polar Programs,
U.S. National Science Foundation-Physics Division,
U.S. National Science Foundation-EPSCoR,
Wisconsin Alumni Research Foundation,
Center for High Throughput Computing (CHTC) at the University of Wisconsin{\textendash}Madison,
Open Science Grid (OSG),
Advanced Cyberinfrastructure Coordination Ecosystem: Services {\&} Support (ACCESS),
Frontera computing project at the Texas Advanced Computing Center,
U.S. Department of Energy-National Energy Research Scientific Computing Center,
Particle astrophysics research computing center at the University of Maryland,
Institute for Cyber-Enabled Research at Michigan State University,
and Astroparticle physics computational facility at Marquette University;
Belgium {\textendash} Funds for Scientific Research (FRS-FNRS and FWO),
FWO Odysseus and Big Science programmes,
and Belgian Federal Science Policy Office (Belspo);
Germany {\textendash} Bundesministerium f{\"u}r Bildung und Forschung (BMBF),
Deutsche Forschungsgemeinschaft (DFG),
Helmholtz Alliance for Astroparticle Physics (HAP),
Initiative and Networking Fund of the Helmholtz Association,
Deutsches Elektronen Synchrotron (DESY),
and High Performance Computing cluster of the RWTH Aachen;
Sweden {\textendash} Swedish Research Council,
Swedish Polar Research Secretariat,
Swedish National Infrastructure for Computing (SNIC),
and Knut and Alice Wallenberg Foundation;
European Union {\textendash} EGI Advanced Computing for research;
Australia {\textendash} Australian Research Council;
Canada {\textendash} Natural Sciences and Engineering Research Council of Canada,
Calcul Qu{\'e}bec, Compute Ontario, Canada Foundation for Innovation, WestGrid, and Compute Canada;
Denmark {\textendash} Villum Fonden, Carlsberg Foundation, and European Commission;
New Zealand {\textendash} Marsden Fund;
Japan {\textendash} Japan Society for Promotion of Science (JSPS)
and Institute for Global Prominent Research (IGPR) of Chiba University;
Korea {\textendash} National Research Foundation of Korea (NRF);
Switzerland {\textendash} Swiss National Science Foundation (SNSF);
United Kingdom {\textendash} Department of Physics, University of Oxford.

\end{document}